\documentclass[pre,twocolumn,amsmath,amssymb,floatfix,showpacs]{revtex4}

\usepackage{graphicx}% Include figure files

\begin{document}

\title{Contextual analysis framework for bursty dynamics} % Force line breaks with \\
\author{Hang-Hyun Jo}
\email{hang-hyun.jo@aalto.fi}
\affiliation{Department of Biomedical Engineering and Computational Science, Aalto University School of Science, P.O. Box 12200, Finland}
\author{Raj Kumar Pan}
\affiliation{Department of Biomedical Engineering and Computational Science, Aalto University School of Science, P.O. Box 12200, Finland}
\author{Juan I. Perotti}
\affiliation{Department of Biomedical Engineering and Computational Science, Aalto University School of Science, P.O. Box 12200, Finland}
\author{Kimmo Kaski}
\affiliation{Department of Biomedical Engineering and Computational Science, Aalto University School of Science, P.O. Box 12200, Finland}
%\affiliation{BECS, Aalto University School of Science, P.O. Box 12200, Finland}

\date{\today}% It is always \today, today,

\begin{abstract}
       To understand the origin of bursty dynamics in natural and social processes we provide a general analysis framework, in which the temporal process is decomposed into sub-processes and then the bursts in sub-processes, called contextual bursts, are combined to collective bursts in the original process. For the combination of sub-processes, it is required to consider the distribution of different contexts over the original process. Based on minimal assumptions for inter-event time statistics, we present a theoretical analysis for the relationship between contextual and collective inter-event time distributions. Our analysis framework helps to exploit contextual information available in decomposable bursty dynamics.
\end{abstract}

% 89.75.Da Scaling phenomena in complex systems
% 05.40.-a Random processes
% 89.20.-a Interdisciplinary applications of physics

%\keywords{SOC, directed sandpile, Abelian symmetry, stochasticity, universality class}
%Use showkeys class option if keyword %display desired

\maketitle

\section{Introduction}

In a wide range of natural and social phenomena, inhomogeneous or non-Poissonian temporal processes have been observed. They are described in terms of $1/f$ noise~\cite{Bak1987,Bak1996} or in terms of bursts that are rapidly occurring events within short time-periods alternating with long periods of low activity~\cite{Barabasi2005,Goh2008,Karsai2012a}. In studies of inhomogeneous temporal processes one finds a unified scaling law for the inter-occurrence time of earthquakes~\cite{Bak2002,Corral2004,Davidsen2013}, $1/f$ frequency scaling and power-law for inter-spike interval distributions in neuronal activities~\cite{Bedard2006,Tsubo2012}, and heavy-tailed inter-event time distributions in human task execution and communication patterns~\cite{Barabasi2005,Vazquez2006,Harder2006,Castellano2009}. The origin of these temporal inhomogeneities has been extensively investigated in terms of self-organized criticality (SOC)~\cite{Bak1996,Paczuski2005}, where temporal inhomogeneities are a consequence of self-similar structure in temporal patterns. On the other hand for bursts other mechanisms have also been suggested, such as memory effects~\cite{Karsai2012a} and inhomogeneous Poisson process with time-varying event rate~\cite{Malmgren2008}.

For more comprehensive understanding of bursty behaviour, let us consider a temporal process that can be decomposed into sub-processes. In other words, a set of events with timings comprises events of different contexts, where each context corresponds to each sub-process. For example, communication events of an individual could be classified as being either family-related or work-related according to the communication partner or content. Then understanding the contextual bursts for events of the same context can give us more detailed insight into collective bursts for all kinds of events. However, the effect of context on bursts has been largely ignored except for a few recent works on human dynamics~\cite{Karsai2012b,Song2012}. In order to relate contextual bursts to collective bursts, the distribution of contexts over the original process must be considered in terms of the ordinal time-frame, where the real timings of events are replaced by their orders in the original event sequence. The ordinal time-frame is useful when the order of events is more crucial for the process than their real timings or when the real timings are not available, like the sequence of words in the text~\cite{Altmann2009}. In addition, the origin of bursts can be explored more explicitly as the effect of any intrinsic temporal patterns, such as circadian and weekly cycles of humans~\cite{Jo2012}, is excluded. Moreover, the human bursty dynamics has often been modelled in terms of the ordinal time-frame by ignoring the real time-frame to some extent~\cite{Barabasi2005,Vazquez2006,Min2009,Jo2011,Jo2012c}. Hence, understanding the relation between contextual bursts in real and ordinal time-frames is essential for bridging the gap between the models and reality. 

In this paper, we provide a general framework for analyzing decomposable bursty dynamics in terms of context and time-frame, by studying a minimal model with uncorrelated inter-event times. Interestingly, the main part of our model can be translated into the broad class of mass transport models~\cite{Majumdar2005,Evans2006}, although they emerged from totally different backgrounds. We find that the statistical properties of contextual bursts in real time-frame can be dominated by either collective bursts or contextual bursts in ordinal time-frame, or be characterized by both. We also show that the real and ordinal time-frames are related successively by means of de-seasoning such that the real time-frame is dilated (contracted) for the moment of high (low) activity~\cite{Jo2012}.

The paper is organized as follows. In Sec.~\ref{sect:model}, we devise and analyze the model with uncorrelated inter-event times to investigate the relationship between inter-event time distributions for collective and contextual bursts in real and ordinal time-frames. In Sec.~\ref{sect:deseason}, we apply the de-seasoning method to successively relate the real and ordinal time-frames. Finally, we summarize the results in Sec.~\ref{sect:summary}.

\begin{figure}[!t]
    \includegraphics[width=\columnwidth]{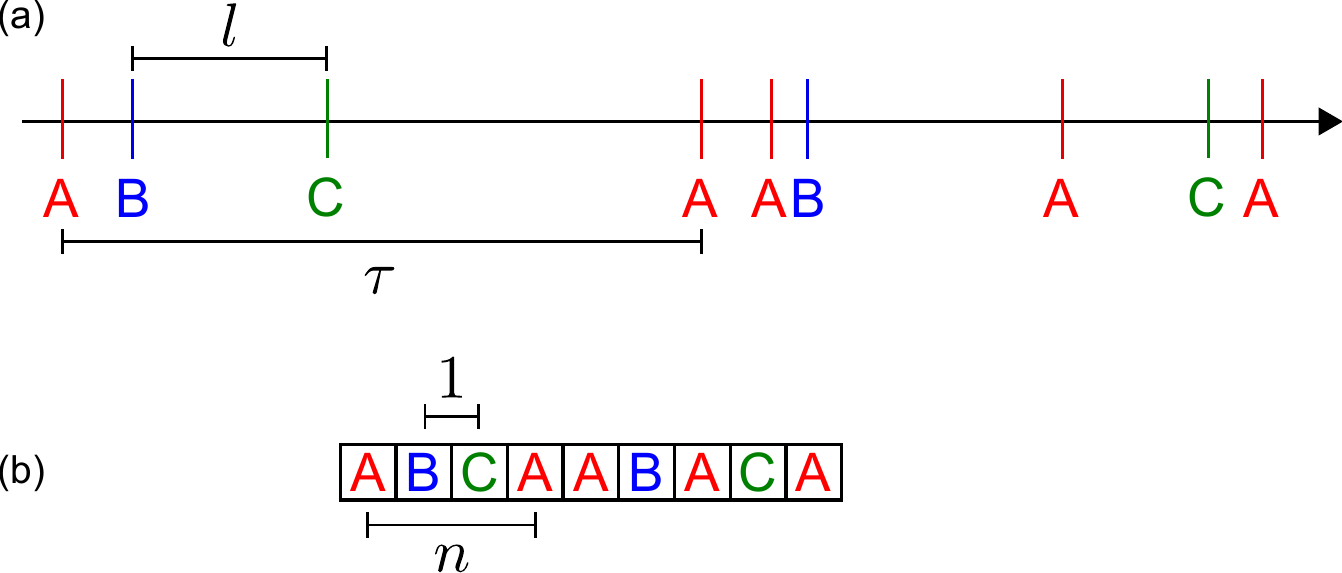}
    \caption{An example of event series with various contexts A, B, and C, presented in real time-frame (a) and in ordinal time-frame (b). $l$ and $\tau$ represent the collective and contextual real inter-event times, respectively. $n$ represents the contextual ordinal inter-event time, while every collective ordinal inter-event time is trivially $1$.}
    \label{fig:timeseries}
\end{figure}

\section{Model}
\label{sect:model}

Let us now introduce an uncorrelated inter-event time model. We denote the collective inter-event time by $l$, whereas contextual inter-event times in real and ordinal time-frames are denoted by $\tau$ and $n$, respectively, see Fig.~\ref{fig:timeseries}. Their corresponding distributions are written by $P(l)$, $P(\tau)$, and $P(n)$. In general, the contextual real inter-event time is obtained by the sum of consecutive collective inter-event times: $\tau=\sum_{i=1}^n l_i$. By means of this relation, the three inter-event time distributions are interrelated as follows:
\begin{eqnarray}
    P(\tau)&=&\sum_{n=1}^\infty P(n) F_n(\tau),\label{eq:Preal}\\
    F_n(\tau)&\equiv &\prod_{i=1}^n \int_{l_0}^\infty dl_i P(l_i) \delta\left(\tau-\sum_{i=1}^n l_i\right).
    \label{eq:Fn}
\end{eqnarray}
Here $F_n(\tau)$ is the probability of obtaining $\tau$ as the sum of $n$ $l$s, each of which is independently drawn from the same distribution $P(l)$. Since only one event can occur at a time in our setup, $l$ must have a positive lower bound, $l_0>0$. When the variance or tail of $P(l)$ is small, one can approximate $\tau=\sum_{i=1}^n l_i \approx n\langle l\rangle$ for sufficiently large $n$, where $\langle \cdot\rangle$ denotes an average. This leads to the trivial solution $P(\tau)\approx P(n)$, implying irrelevance of time-frame. As the general case, we consider the heavy-tailed distribution, $P(l)\propto l^{-\alpha}$ with $\alpha>1$. The distribution of $P(n)$ is closely related to the context distribution over the event sequence. For the case with very few contexts, as $n$ is mostly $1$, i.e. $\tau=l$, we obtain $P(\tau)\approx P(l)$, implying irrelevance of context. In general we assume that the contexts are unevenly distributed over the event sequence by considering $P(n)\propto n^{-\beta}$ with $\beta>1$. Then we find that $P(\tau)$ shows an asymptotic power-law behavior, $\tau^{-\alpha'}$.

\subsection{Main results}

\begin{figure}[!t]
    \includegraphics[width=.5\columnwidth]{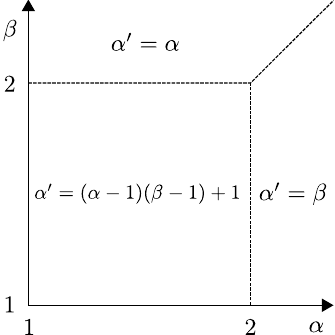}
    \caption{Phase diagram summarizing the relation between heavy-tailed distributions of $l$, $\tau$, and $n$, in terms of their corresponding power-law exponents $\alpha$, $\alpha'$, and $\beta$. Contextual bursts in real time-frame are dominated by contextual bursts in ordinal time-frame if $\alpha>2$ and $\alpha>\beta$, by collective bursts if $\beta>2$ and $\beta>\alpha$, or otherwise characterized by both kinds of bursts.}
    \label{fig:result}
\end{figure}

In Fig.~\ref{fig:result}, we depict the main results. Both collective bursts and contextual bursts in ordinal time-frame are generally expected to affect contextual bursts in real time-frame. This is the case only when both kinds of bursts are sufficiently strong, i.e. $\alpha'=(\alpha-1)(\beta-1)+1$ for $\alpha<2$ and $\beta<2$. This scaling relation can be understood by the identity $P(\tau)d\tau=P(n)dn$ with the fact that $\tau=\sum_{i=1}^n l_i$ is dominated by $\max\{l_i\}$ that is proportional to $n^{1/(\alpha-1)}$~\cite{Braunstein2003}. On the other hand, when $\alpha>2$ and $\alpha>\beta$, it turns out that the same power-law exponent characterizes contextual bursts in both time-frames, i.e. $\alpha'=\beta$. This implies that the time-frame is not relevant to contextual bursts. Finally, when $\beta>2$ and $\beta>\alpha$, we find $\alpha'=\alpha$, implying that the context distribution over the event sequence is not relevant to bursts in real time-frame. 

\subsection{Analysis}

For analysis, we change variables by $m_i\equiv l_i-l_0$ and $M\equiv \tau-nl_0$ to rewrite $Z_n(M)\equiv F_n(M+nl_0)$ and $f(m_i)\equiv P(m_i+l_0)$:
\begin{equation}
    Z_n(M)=\prod_{i=1}^n \int_0^\infty dm_i f(m_i)\delta\left(M-\sum_{i=1}^n m_i\right).
\end{equation}
This is exactly the ``canonical partition function'' for mass transport models and its analytical solution for $f(m)\simeq Am^{-\alpha}$ has been extensively studied~\cite{Majumdar2005,Evans2006}. 

For $1<\alpha<2$, $Z_n(M)$ follows a scaling form as~\cite{Evans2006}
\begin{equation*}
    Z_n(M)\simeq\left\{\begin{tabular}{ll}
	$n^{-\nu}g_1^-(Mn^{-\nu})$ & if $M<n$\\
	$n^{-\nu}g_1^+(Mn^{-\nu})$ & if $M>n$\\
    \end{tabular}\right.,
\end{equation*}
with $\nu=\frac{1}{\alpha-1}$ and the scaling functions are
\begin{eqnarray}
    g_1^-(x)&=&a x^{-\frac{3-\alpha}{2(2-\alpha)}}\exp\left(-b x^{-\frac{\alpha-1}{2-\alpha}}\right),\label{eq:g1}\\
    g_1^+(x)&=&c x^{-\alpha},
\end{eqnarray}
where $a$, $b$, and $c$ are constants depending on $\alpha$ and $A$~\footnote{Since the saddle point approximation for calculating $g_1^-(x)$, i.e. Eq.~(118) in~\cite{Evans2006}, is not valid for large $x$, we use the alternative approximation to obtain $g_1^+(x)$ as described in Appendix A.3.1 in~\cite{Evans2006}.}. After plugging this scaling form into Eq.~(\ref{eq:Preal}), we perform the summation over $n$ with the upper bound of $\frac{\tau}{l_0}$ due to $M\geq 0$. Then, we get
\begin{eqnarray}
    P(\tau)&=&\sum_{n=1}^{\tau/l_0} P(n) Z_n(\tau-nl_0)\nonumber\\
    &\propto &\int_1^{n_\times} n^{-\beta-\nu} g_1^+[(\tau-nl_0)n^{-\nu}]dn\nonumber\\
    & &+\int_{n_\times}^{\tau/l_0} n^{-\beta-\nu} g_1^-[(\tau-nl_0)n^{-\nu}]dn\nonumber\\
    &\propto &\tau^{-\alpha_c}\left[\int_0^{x_\times} x^{\alpha_c-1}g_1^-(x)dx+ \int_{x_\times}^{\tau} x^{\alpha_c-1}g_1^+(x)dx\right]\nonumber\\\label{eq:Preal0}
\end{eqnarray}
with $\alpha_c\equiv(\alpha-1)(\beta-1)+1$ and crossovers $n_\times$ and $x_\times=(\tau-n_\times l_0)n_\times^{-\nu}$. For derivation, $(\tau-nl_0)n^{-\nu}$ has been replaced by $x$ and then approximated as $x\approx\tau n^{-\nu}$. While the first term in the parenthesis is independent of $\tau$, the second term is obtained as $\tau^{\alpha_c-\alpha}- x_\times^{\alpha_c-\alpha}$, leading to
\begin{equation}
    P(\tau)\propto c_1\tau^{-\alpha_c}+c_2\tau^{-\alpha}
    \label{eq:Preal1}
\end{equation}
with coefficients $c_1$ and $c_2$. Thus, we obtain 
\begin{equation}
    \alpha'=\min\{\alpha_c,\alpha\}\ \textrm{if}\ 1<\alpha<2.
    \label{eq:alpha1}
\end{equation}
The condition for $\alpha_c=\alpha$ is $\beta=2$, when the second term in Eq.~(\ref{eq:Preal0}) gives the logarithmic correction as $\ln\tau$. That is, if the tail of $P(n)$ is sufficiently small, $\alpha'=\alpha$ is obtained, implying that contextual bursts in real time-frame is determined only by collective bursts. In any case, we get $\alpha'<\beta$, implying that contextual bursts in real time-frame are stronger than those in ordinal time-frame due to large fluctuations of collective inter-event times.

\begin{figure}[!t]
    \includegraphics[width=\columnwidth]{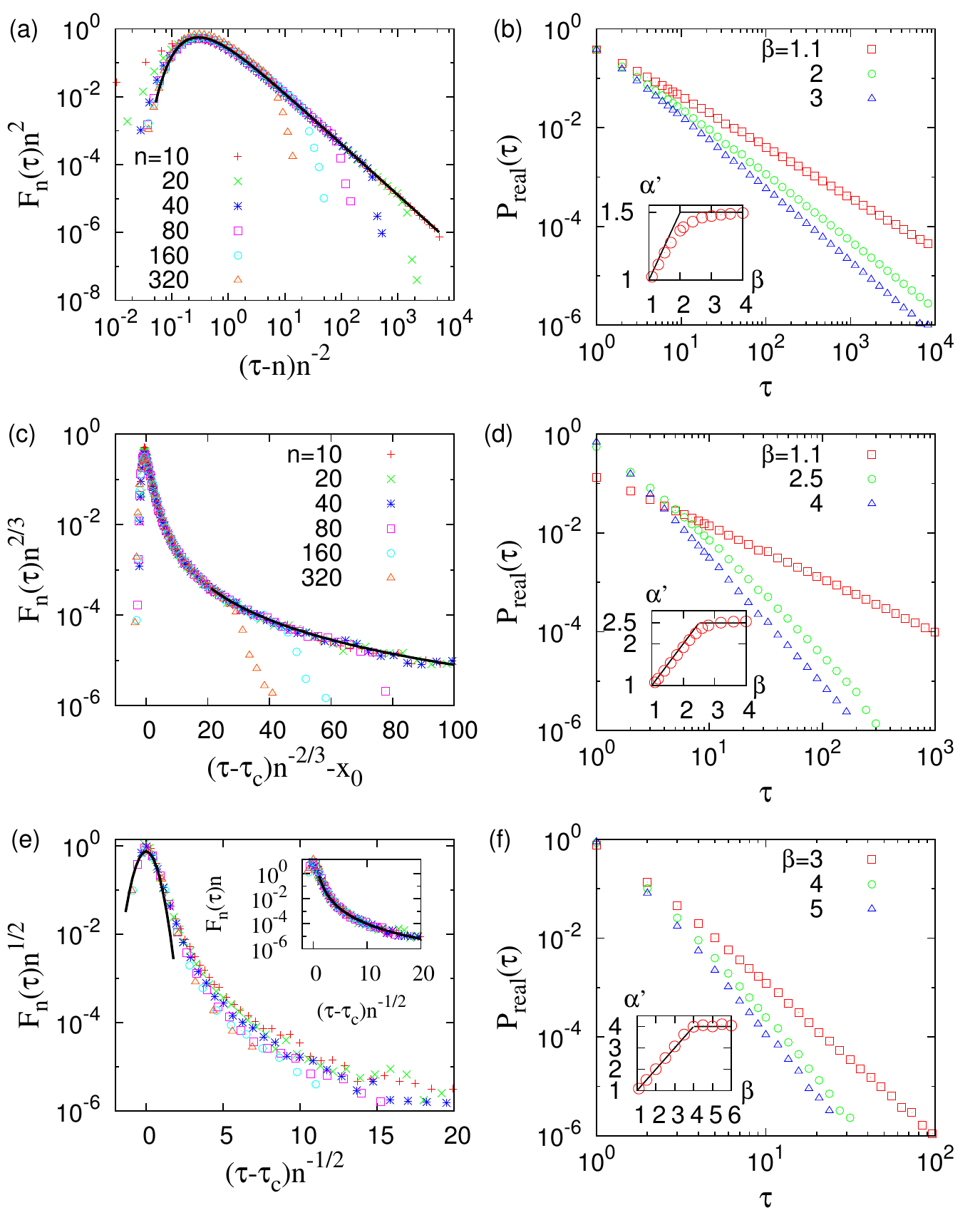}
    \caption{Numerical results of the model for $\alpha=\frac{3}{2}$ (top), $\frac{5}{2}$ (middle), and $4$ (bottom), all with $l_0=1$. (a) Numerical results of $F_n(\tau)$ for different values of $n$ collapse into one curve, i.e. $g_1^-(x)$ for $x<x_\times$ and $g_1^+(x)$ for $x>x_\times$, drawn with black curve. (b) The power-law exponent $\alpha'$ is estimated from $P(\tau)\sim \tau^{-\alpha'}$ for each $\beta$ to be compared with the scaling relation $\alpha'=\min\{\frac{\beta+1}{2},\frac{3}{2}\}$, denoted by black line in the inset. (c-f) Scaling collapse of $F_n(\tau)$ and the estimated values of $\alpha'$ comparable to $\alpha'=\min\{\alpha,\beta\}$ support our analysis for $\alpha>2$. Here $\tau_c$ for each $n$ is determined to maximize $F_n(\tau)$. In (c) $x_0\approx 0.659$ is used. In the inset of (e), tail parts are collapsed by $F_n(\tau)n$ versus $(\tau-\tau_c)n^{-1/2}$.}
    \label{fig:numerics}
\end{figure}

Figure~\ref{fig:numerics}~(a,b) shows that our analysis is confirmed by the numerical simulations (to be described later) for $\alpha=\frac{3}{2}$ and $l_0=1$. We find that the numerically obtained $F_n(\tau)$ for different $n$s collapse into one curve corresponding to $g_1^-(x)$ for $x<x_\times$ and $g_1^+(x)$ for $x>x_\times$. Then, based on the simple scaling form, $P(\tau)\sim\tau^{-\alpha'}$, we estimate the value of $\alpha'$, which shows slight discrepancy from the analytic result in Eq.~(\ref{eq:alpha1}), due to the correction term in Eq.~(\ref{eq:Preal1}).

Next, for more realistic considerations like finite-size data, we discuss the effect of cutoff by assuming that $P(n)\propto n^{-\beta} h(\frac{n}{n_c})$ with a cutoff function $h(x)$. Let us consider the case of steep cutoff, i.e. $h(x)=1$ for $x\leq 1$ and $0$ for $x>1$. If $\tau$ is sufficiently large as $\tau>x_\times n_c^\nu$, we obtain the asymptotic result, $\alpha'=\alpha$, implying that $\alpha'$ is determined only by $\alpha$. In case of exponential cutoff, $h(x)=e^{-x}$, the same result, $\alpha'=\alpha$, is also confirmed by numerical simulations (not shown).

For $2<\alpha<3$, $Z_n(M)$ is a function of $M-M_c$ with the ``critical point'' $M_c=n\langle m\rangle$, at which the condensation transition occurs~\cite{Evans2006}. Thus, we separate the subcritical and supercritical cases as follows: 
\begin{equation*}
    Z_n(M)\simeq\left\{\begin{tabular}{ll}
	$n^{-\nu}g_2^-[(M_c-M)n^{-\nu}]$ & if $M<M_c$\\
	$n^{-\nu}g_2^+[(M-M_c)n^{-\nu}]$ & if $M>M_c$\\
    \end{tabular}\right.,
\end{equation*}
where $\nu=\frac{1}{\alpha-1}$. $g_2^-(x)$ has the same form as $g_1^-(x)$ in Eq.~(\ref{eq:g1}), and $g_2^+(x)\propto x^{-\alpha}$. Since $M-M_c=\tau-n(l_0+\langle m\rangle)=\tau-n\langle l\rangle$, we split the summation over $n$ in Eq.~(\ref{eq:Preal}) at $\frac{\tau}{\langle l\rangle}$ as follows:
\begin{eqnarray*}
    P(\tau) &\propto &\int_1^{\tau/\langle l\rangle} n^{-\beta-\nu} g_2^+[(\tau-n\langle l\rangle)n^{-\nu}]dn\\
    & &+\int_{\tau/\langle l\rangle}^{\tau/l_0} n^{-\beta-\nu} g_2^-[(n\langle l\rangle -\tau)n^{-\nu}]dn
\end{eqnarray*}
Similarly to the calculation for $g_1^+(x)$ in Eq.~(\ref{eq:Preal0}), the first term gives the form of $\tau^{-\alpha_c}+\tau^{-\alpha}$. For the second term, we assume that $g_2^-(x)\approx\delta(x-x_0)$, where the location of peak $x_0>0$ is defined by ${g_2^-}'(x_0)=0$~\footnote{Explicitly, $x_0=[\frac{2(\alpha-2)}{3-\alpha}]^{\frac{\alpha-2}{\alpha-1}} [(\alpha-1)A\Gamma(1-\alpha)]^{-\frac{1}{\alpha-1}}$ with $A^{-1}=\int_0^\infty m^{-\alpha}dm$. Since $A$ is not defined due to the singularity at $m=0$, we use $A\approx\zeta(\alpha)^{-1}$ for the discrete case. For $\alpha=\frac{5}{2}$, $x_0\approx 0.659$, which is used in Fig.~\ref{fig:numerics}~(c).}. Since the root of the equation $\frac{n\langle l\rangle-\tau}{n^\nu}-x_0=0$ is in the range of $(\frac{\tau}{\langle l\rangle}, \frac{\tau}{l_0})$, one finds $\tau^{-\beta}$, leading to $\alpha'=\min\{\alpha_c,\alpha,\beta\}$. Knowing that $\beta<\alpha_c$ for $\alpha>2$, we find
\begin{equation}
    \alpha'=\min\{\alpha,\beta\}\ \textrm{if}\ 2<\alpha<3.
    \label{eq:alpha2}
\end{equation}
In other words, collective bursts and contextual bursts in ordinal time-frame compete for contextual bursts in real time-frame. In particular, the result $\alpha'=\beta$ for $\beta<\alpha$ implies that the approximation $\tau=\sum_{i=1}^n l_i\approx n\langle l\rangle$ is still valid even when $\langle l^2\rangle$ diverges. It is because only the subcritical part of $Z_n(M)$, where the fluctuation of $l$ is negligible, contributes to $P(\tau)$. 

For $\alpha>3$, $Z_n(M)$ can be written by means of central and peripheral scaling functions as~\cite{Evans2006}
\begin{equation*}
    Z_n(M)\simeq\left\{\begin{tabular}{ll}
	$n^{-\nu}g_3^<[(M-M_c)n^{-\nu}]$ & if $|M-M_c|\lesssim \mathcal{O}(n^{\frac{2}{3}})$\\
	$n^{-\mu}g_3^>[(M-M_c)n^{-\nu}]$ & if $M-M_c\gtrsim \mathcal{O}(n)$\\
    \end{tabular}\right.,
\end{equation*}
where $\nu=\frac{1}{2}$ and $\mu=\frac{\alpha}{2}-1$. The central scaling function is 
\begin{equation}
    g_3^<(x)=\frac{1}{\sqrt{2\pi \Delta^2}}\exp\left(-\frac{x^2}{2\Delta^2}\right),
\end{equation}
where $\Delta^2=\langle m^2\rangle-\langle m\rangle^2$. The peripheral scaling function $g_3^>(x)$ is the same as $g_2^+(x)$. By assuming that $g_3^<(x)\approx\delta(x)$, we obtain $\alpha'=\min\{\alpha,\beta\}$. Our analysis is confirmed by the numerical results as shown in Fig.~\ref{fig:numerics}~(c-f). Finally, all analytical results are summarized as
\begin{equation}
    \alpha'=\min\{(\alpha-1)(\beta-1)+1,\alpha,\beta\}
    \label{eq:alphaAll}
\end{equation}
and depicted in Fig.~\ref{fig:result}. 

In our numerical methods, $l$ is considered to be an integer starting from $l_0=1$, so is $\tau$. We prepare a set of collective real inter-event times as $L=\{1,\cdots,2,\cdots,l_{\rm max}\}$, where the number of $l$ is proportional to $P(l)= Al^{-\alpha}$ with $A^{-1}=\sum_{l=1}^{l_{\rm max}}l^{-\alpha}$. Here $l_{\rm max}$ is determined under the condition $A^{-1}>0.999~\zeta(\alpha)$. When $n$ is given, we randomly select $n$ elements from the set $L$ and get the sum of them as $\tau=\sum_{i=1}^n l_i$, which is repeated up to $10^9$ times to make the distribution $F_n(\tau)$. By plugging these $F_n(\tau)$ into Eq.~(\ref{eq:Preal}) together with $P(n)$, we numerically obtain $P(\tau)$.

\section{De-seasoning method}
\label{sect:deseason}

Although real and ordinal time-frames are qualitatively different, we can successively relate them in terms of the de-seasoning method~\cite{Jo2012}. In order to de-season intrinsic cyclic activity, denoted by $\rho(t)$, the real time-frame is dilated (contracted) for the moment of high (low) activity. Let us denote the number of events at time $t$ by $a(t)$, being either $0$ or $1$ in our setup. Given the de-seasoning period $T$, the event rate reads 
\begin{equation}
    \rho(t)=\frac{T}{s}\sum_k a(t+kT)
\end{equation}
with the total number of events $s$ and by the normalization, $\frac{1}{T}\int_0^T\rho(t)dt=1$. The de-seasoned time $t^*$ is defined by means of $\rho^*(t^*)dt^*=\rho(t)dt$ with $\rho^*(t^*)=1$, which implies no cyclic patterns in the de-seasoned time-frame. Correspondingly, the de-seasoned inter-event time between event timings $t_1$ and $t_2$ is defined by $\tau^*=\int_{t_1}^{t_2}\rho(t')dt'$. As the minimum of de-seasoned inter-event time is $\frac{T}{s}$, the domain of de-seasoned inter-event time distribution becomes smaller for the larger de-seasoning period. This means that the de-seasoning generically leads to less bursty behavior. 

When the time series is fully de-seasoned, i.e. $T=T_f$ with the entire period $T_f$, we get $\rho(t)=\frac{T_f}{s}a(t)\equiv l_0^* a(t)$. Since every collective de-seasoned inter-event time is $l_0^*$, the contextual de-seasoned inter-event time must be a multiple of $l_0^*$, such that $\tau^*=nl_0^*$. Here $n$ denotes the contextual ordinal inter-event time. Conclusively, all temporal properties in the fully de-seasoned real time-frame should be identical to those in the ordinal time-frame. This in turn leads to an interesting question whether contextual bursts in real and ordinal time-frames can also be successively related. 

\section{Summary}
\label{sect:summary}

In summary, we have provided a general framework for analyzing decomposable bursty dynamics in terms of context and time-frame, by studying an uncorrelated inter-event time model. We derived asymptotic relationships between the collective bursts and contextual bursts in real and ordinal time-frames. We found that the contextual bursts in real time-frame can be dominated by either collective bursts or contextual bursts in ordinal time-frame, or be characterized by both kinds of bursts. This implies that collective bursts may have different origins. In particular, the (in)difference between the contextual bursts in real and ordinal time-frames is important to relate models in ordinal time-frame with the real systems. Our framework of decomposing a temporal process into sub-processes and combining them after understanding each sub-process helps us to investigate complex systems showing temporal inhomogeneities like $1/f$ noise or bursts, in more detail. Although temporal inhomogeneities could be understood to some extent only by inter-event time distributions, it is important to extend our minimal model to take various correlations and memory effects into account.

\begin{acknowledgments}
Financial support by Aalto University postdoctoral program (HJ), by the Academy of Finland, the Finnish Center of Excellence programme 2006-2011, project no. 129670 (RKP, KK) is gratefully acknowledged.
\end{acknowledgments}

\bibliography{h2jo-timeframe}
\end{document}